\documentclass{PoS}

\usepackage{amsmath}
\usepackage{amsfonts}
\usepackage{amssymb}
\usepackage{graphicx}
\usepackage{fontenc}
\usepackage{times}
\usepackage{mathptmx}

\newcommand{\Tr}{\mathrm{Tr}}

\newcommand{\VEV}[1]{\left\langle #1 \right\rangle}

\newcommand{\badj}{\beta_{\rm adj}}
\newcommand{\bfund}{\beta_{\rm fund}}

\title{Scaling properties of SU(2) gauge theory with mixed
  fundamental-adjoint action }

\ShortTitle{SU(2) gauge theory with fundamental-adjoint action }

\author{\speaker{Enrico Rinaldi}\thanks{The author is supported by a
    SUPA prize studenship and a JSPS short-term fellowship.}\\
    SUPA and The Tait Institute, School of Physics and Astronomy,
    University of Edinburgh\\
    Edinburgh, EH9 3JZ, UK\\
    and
    Kobayashi-Maskawa Institute, Nagoya University,
    Nagoya, 464-8602, Japan\\
    E-mail: \email{e.rinaldi@sms.ed.ac.uk}}

\author{Giuseppe Lacagnina\\
        E-mail: \email{giuseppe.lacagnina@gmail.com}}

\author{Biagio Lucini\\
        College of Science, Swansea University, Swansea, SA2 8PP, UK\\
        E-mail: \email{B.Lucini@swansea.ac.uk}}

\author{Agostino Patella\\
        School of Computing and Mathematics, Plymouth, PL4 8AA, UK\\
        and PH-TH, CERN, CH-1211 Geneva 23, Switzerland\\
        E-mail: \email{agostino.patella@plymouth.ac.uk}}

\author{Antonio Rago\\
        School of Computing and Mathematics, Plymouth, PL4 8AA, UK\\
        E-mail: \email{antonio.rago@plymouth.ac.uk}}

\abstract{We study the phase diagram of the SU(2) lattice gauge theory
  with fundamental--adjoint Wilson plaquette action. We confirm the
  presence of a first order bulk phase transition and we estimate the
  location of its end--point in the bare parameter space. If this point
  is second order, the theory is one of the simplest realizations of a
  lattice gauge theory admitting a continuum limit at finite
  bare couplings. All the relevant gauge observables are monitored in
  the vicinity of the fixed point with very good control over
  finite--size effects. The scaling properties of the low--lying
  glueball spectrum are studied while approaching the end--point in a
  controlled manner.\\
  [0.25cm]
  \rightline{CERN-PH-TH/2012-295}
}

\FullConference{The 30 International Symposium on Lattice Field Theory
  - Lattice 2012,\\
  June 24-29, 2012\\
  Cairns, Australia}

\begin{document}

\section{Introduction}
\label{sec:introduction}

The simplest lattice discretization of the SU($N_c$) Yang--Mills
theory is the well--known Wilson plaquette
action~\cite{wilson}. Different discretization of the lattice action
will not change the physics of the continuum limit realised in the
neighborhood of the weakly coupled ultravioled fixed point. However, far from
this continuum limit, different discretizations can lead to the
appearence of second order phase transition points that can mimic a
continuous infrared fixed point for the theory defined by the naive
lattice dicretization. In fact, although a continuum theory can be
defined at any of those points, in principle this theory is not
related to the ultraviolet gaussian fixed point.
%spurious fixed points.
% When studying a given continuum theory, the
% choice of the lattice action is not unique. A different choice of the
% discretized action can lead to different lattice artefatcs; moreover,
% spurious fixed points, unrelated to the continuum physics we are
% interested in, can appear.
\\
One possible extension of the Wilson action includes plaquette terms
in a representation of the gauge group other than the fundamental. For
example the following action includes a term in the adjoint representation
\begin{equation}
  \label{eq:action-lattice}
  S = \bfund \sum_{i,\mu > \nu} \left( 1 - \frac{1}{N_c} \mbox{Re} \ \Tr_F
    \left( U_{\mu \nu}(i) \right) \right) + \badj \sum_{i,\mu > \nu}
  \left( 1 - \frac{1}{N_c^2 - 1} \mbox{Re} \ \Tr_A
    \left( U_{\mu \nu}(i) \right) \right) 
  \ ,
\end{equation}
where $N_c$ is the number of colours and $U_{\mu \nu}(i)$ the plaquette
in the $(\mu,\nu)$--plane from point $i$. The sum over all the points
$i$ is done over the four--dimensional hypercubic lattice
$L^4$. $\Tr_F$ and $\Tr_A$ are,
respectively, the trace defined in the fundamental and
in the adjoint representation of the SU($N_c$) gauge group. They are
related by $\Tr_A (U) = |\Tr_F(U)|^2 - 1$.\\
This fundamental--adjoint plaquette action has been used in the
pioneering work of Ref.~\cite{bhanot} and in several more recent
studies~\cite{gavai} which extensively investigated the structure of the
phase diagram for $N_c=2$. % The SU(3) case was also considered, for
% example in Ref.~\cite{heller}.
Our interest in this model comes from
the recent studies of the conformal window using lattice field theory
techniques. The SU($2$) gauge theory with $2$ adjoint fermions has
been shown to have an infrared conformal fixed point 
by looking at the scaling properties of the mesonic and gluonic
spectrum~\cite{luigi}. In principle, the same features
could be reproduced around a second-order phase transition  point appearing as a
lattice artefact due to the chosen lattice discretization. The
fundamental--adjoint lattice model of Eq.~\eqref{eq:action-lattice}
can be seen as the leading contribution to the
action with adjoint fermions in the heavy bare quark mass
limit. Hence, if the
end--point of the first order phase transition in this model
turns out to be a lattice--induced second order phase transition
point, one needs to investigate how the
results of Ref.~\cite{luigi} would be affected by it. In the following, we
investigate carefully the phase diagram and the spectrum of the lattice model
in the vicinity of the end--point to check whether the infrared
physics resembles the one studied in Ref.~\cite{luigi}. A detailed
description of our study will be the object of a forthcoming
publication~\cite{mypaper}.

\section{Phase diagram}
\label{sec:phase-diagram}

\begin{figure}[ht]
  \centering
  \begin{tabular}[h]{cc}
    \includegraphics[width=0.45\textwidth]{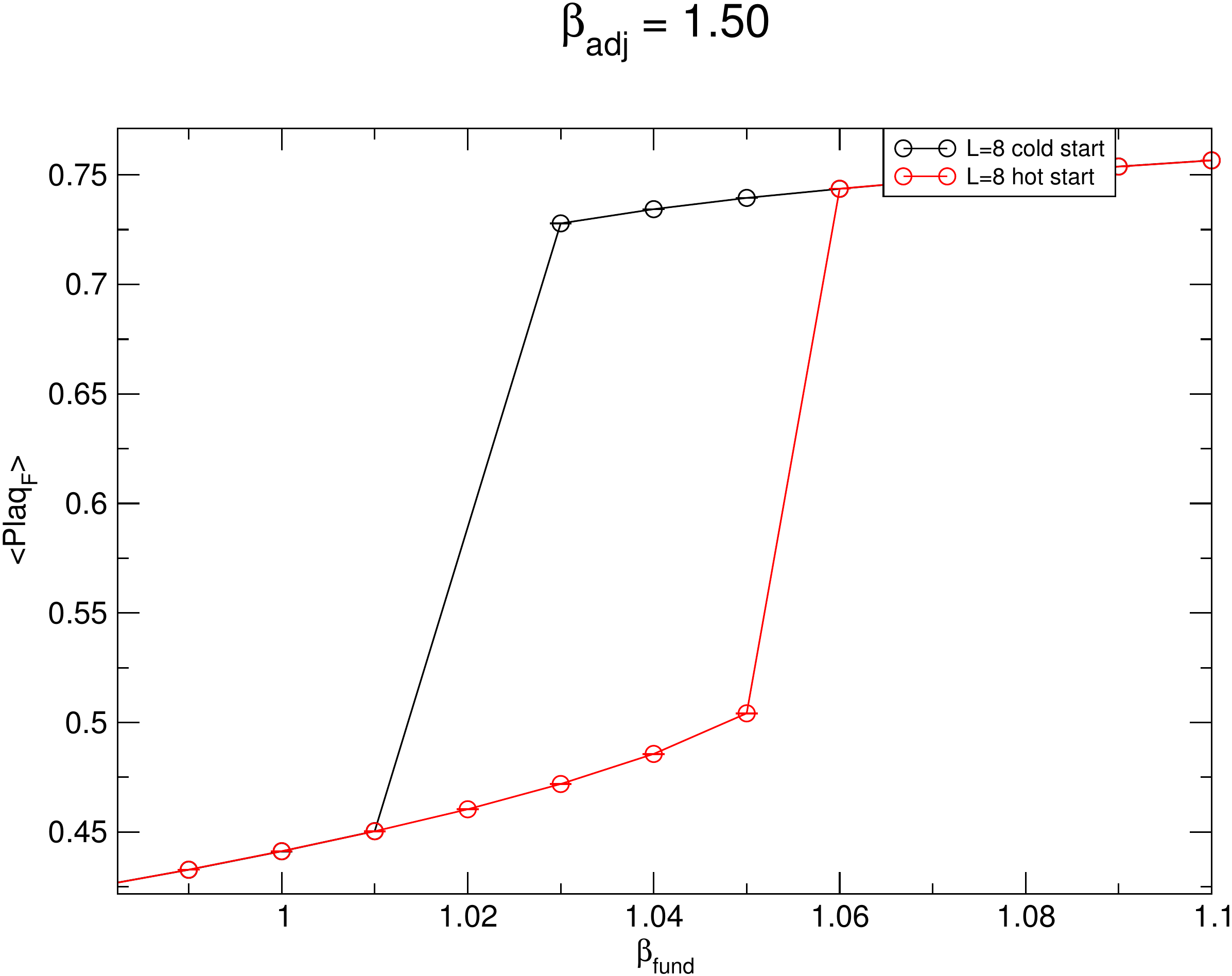} &
    \includegraphics[width=0.45\textwidth]{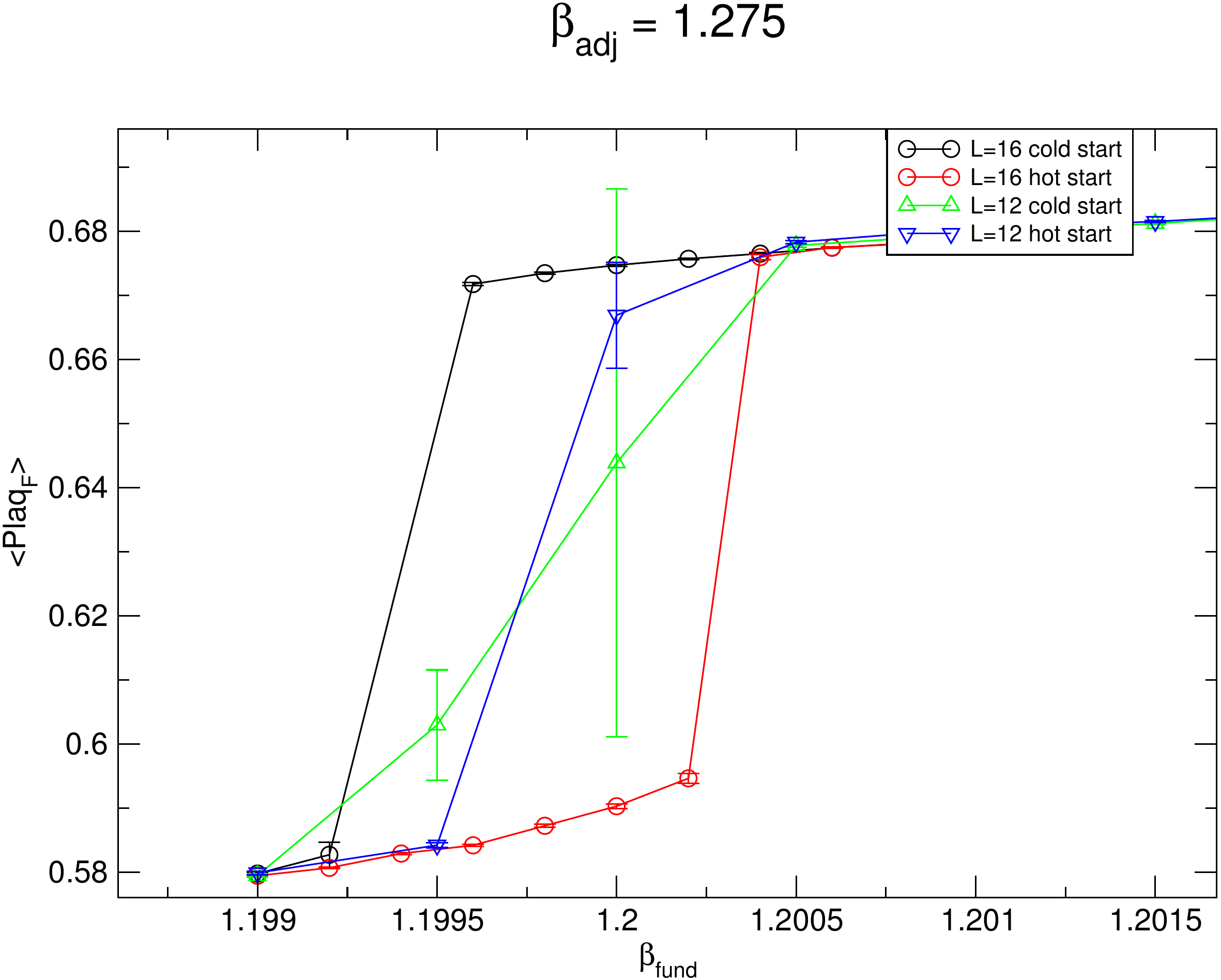} \\
  \end{tabular}
  \caption{Hysteresis cycle of the fundamental plaquette for different
    values of the adjoint
    coupling $\badj$. At large $\badj$ (left), the
    separation between the hysteresis branches is visible on
    relatively small volumes. At smaller $\badj$ (right) bigger volumes
    are necessary to clearly distinguish the first order nature of
    the transition.}
  \label{fig:hysteresis}
\end{figure}

In the two--dimensional plane of the coupling constants
($\bfund$,$\badj$), the theory presents several regions % . On the adjoint
% axis $\bfund=0$ there is a first order phase transition that enters
% the plane at non--zero fundamental coupling. On the other hand, at
% $\badj=\infty$ a first order phase transition enters the plane at
% finite $\badj$, merging together with the aforementioned line
(cfr. Ref.~\cite{bhanot} for a qualitative picture). On the
fundamental axis $\badj=0$ the SU(2) gauge theory with standard Wilson
plaquette action is recovered and this is known to have a crossover
region at $\bfund \approx 2.30$. When the adjoint coupling is turned
on and the second term of Eq.~\eqref{eq:action-lattice} starts
becoming important, the system develops a first order bulk transition
which becomes stronger as $\badj$ increases. We have monitored the
location of this bulk transition line by studying the
expectation value of the fundamental and adjoint plaquettes,
% \begin{equation}
%   \label{eq:plaquette-vev}
%   {\rm Plaq}_F \; = \; \frac{1}{6L^4}\frac{1}{2} \sum_{i,\mu > \nu}
%   \mbox{Re}\Tr_F U_{\mu \nu}(i) \; {\rm ;}
%   \qquad
%   {\rm Plaq}_A \; = \; \frac{1}{6L^4}\frac{1}{3} \sum_{i,\mu > \nu}
%   \mbox{Re}\Tr_A U_{\mu \nu}(i)
%   \ ,
% \end{equation}
and we have also computed the corresponding normalized susceptibilities.
% \begin{equation}
%   \label{eq:plaquette-susc}
%   \frac{\chi_{Pf}}{L^4} \; = \; \VEV{{\rm Plaq}_F^2} - \VEV{{\rm
%       Plaq}_F}^2 \; {\rm ;}
%   \qquad
%   \frac{\chi_{Pa}}{L^4} \; = \; \VEV{{\rm Plaq}_A^2} - \VEV{{\rm Plaq}_A}^2 
%   \ .
% \end{equation}
An example of the hysteresis cycle characteristic of the bulk phase
transition at large $\badj=1.50$ is shown in
Fig.~\ref{fig:hysteresis}(left): on a hypercubic symmetric lattice
of size $8^4$ we clearly distinguish two separate branches for the
fundamental plaquette as $\bfund$ is changed starting from a
random (hot) or unit (cold) gauge configuration. When $\badj$ is
decreased, we note that larger volumes are necessary in
order to correctly identify the presence of the hysteresis loop. For
example, at $\badj=1.275$ a lattice $12^4$ is not large enough for the
system to develop the two vacua of the first order transition, and this
is shown in Fig. \ref{fig:hysteresis}(right).\\

%%%%%%%%%%%%%%
%%%%%%%%%%%%%%

%%%%%%%%%%%%%%
\begin{figure}[ht]
  \centering
  \begin{tabular}[h]{cc}
    \includegraphics[width=0.45\textwidth]{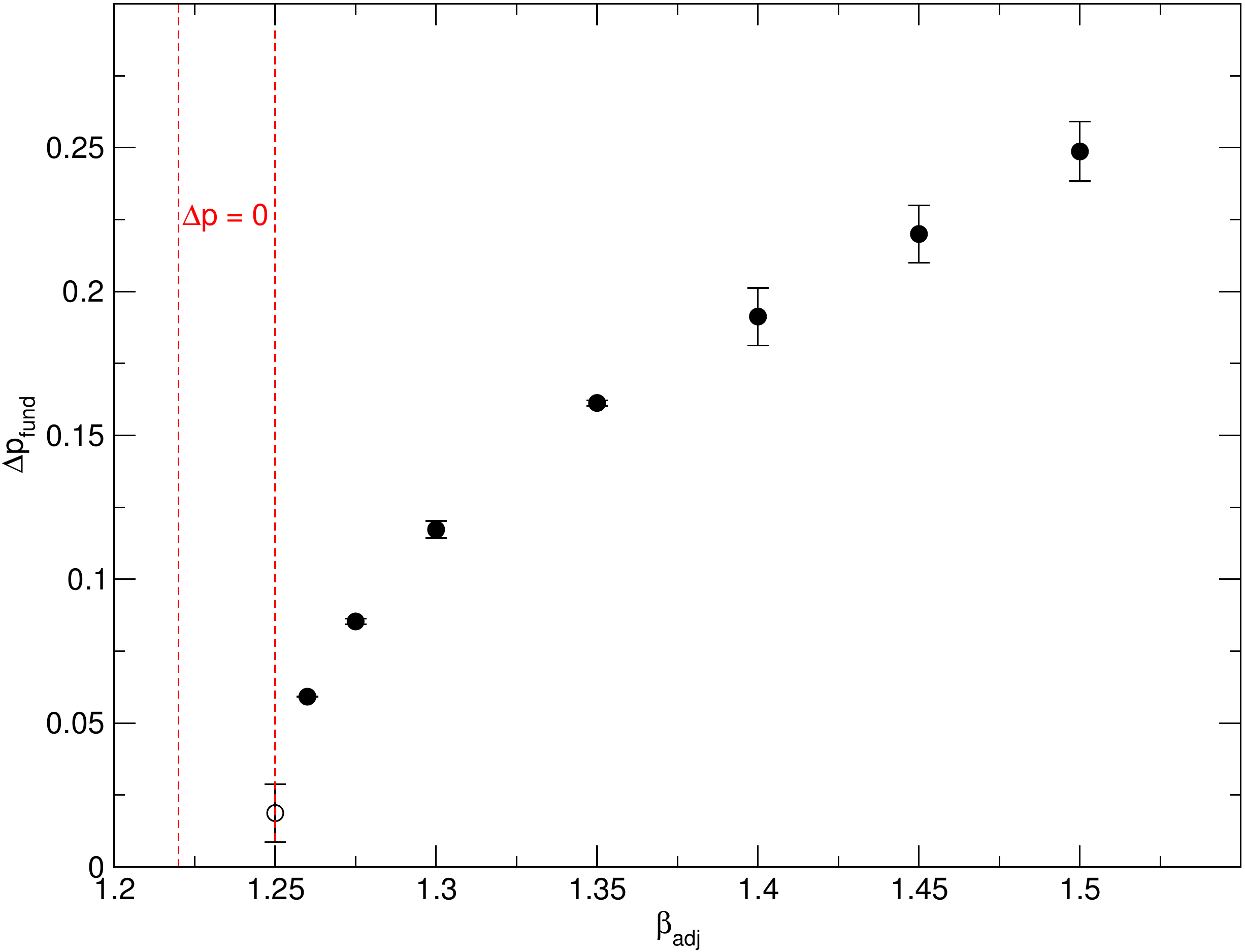} &
  \end{tabular}
  \caption{Fundamental plaquette difference between hot
    and cold start runs at the centre of the hysteresis cycle. Also
    shown is the approximate position of the critical $\badj$ value
    at which this difference is expected to vanish. A consistent
    result is found using the adjoint plaquette.}
  \label{fig:latent-heat}
\end{figure}
%%%%%%%%%%%%%%
By further decreasing $\badj$, the separation between the lower and
the upper branch of the hysteresis shows a clear trend suggesting that
it should vanish at approximately $\badj \lesssim 1.25$. An estimate
of this separation is given by the difference of the plaquette in the
two vacua at the center of the hysteresis loop:
\begin{equation}
  \label{eq:latent-heat}
  \Delta p_{\rm fund} \; = \; \VEV{{\rm Plaq}_{F,1}} -
  \VEV{{\rm Plaq}_{F,2}}
  \ ,
\end{equation}
where the subscripts $1$ and $2$ refer to the disctinct vacua, and a
similar definition holds for $\Delta p_{\rm adj}$. This $\Delta p_{\rm fund}$ is
plotted in Fig.~\ref{fig:latent-heat}, where we always used the smallest
volume where the first order nature of the transition was
manifest. The point at $\badj=1.25$ required a very large $40^4$
lattice for which we currently do not have very good control over the
systematic and statistical uncertainties of the simulation. 
% Since the volume needed at the smallest $\badj$ is as large
% as $40^4$ and autocorrelation times become increasingly longer, our
% estimate for the location of the transition's end--point is affected
% by sizable statistical error.
\\
In the region below the approximate location of the end--point, we have
checked that the transition 
becomes a crossover, signalled by the lack of scaling with the volume
in the fundamental and adjoint plaquette susceptibilities. The height and
the location of peak of the susceptibility is consistent across the
different volumes. % An example for two
% different $\badj$ values is reported in Fig.~\ref{fig:plaquette-susc},
% where in one case lattice sizes up to $28^4$ were used.
The location
of the peak separates a strong coupling region at small $\bfund$ from
a region closer to the weak coupling limit ($\bfund \rightarrow
\infty$).
%%%%%%%%%%%%%%
% \begin{figure}[ht]
%   \centering
%   \begin{tabular}[h]{cc}
%     \includegraphics[width=0.45\textwidth]{FIGS/susc_1-00} &
%     \includegraphics[width=0.45\textwidth]{FIGS/susc_1-16} \\
%   \end{tabular}
%   \caption{Lack of scaling in the fundamental plaquette susceptibility
%   for two values of $\badj < 1.25$.}
%   \label{fig:plaquette-susc}
% \end{figure}
%%%%%%%%%%%%%%

%%%%%%%%%%%%%%
\begin{figure}[ht]
  \centering
  \includegraphics[width=0.50\textwidth]{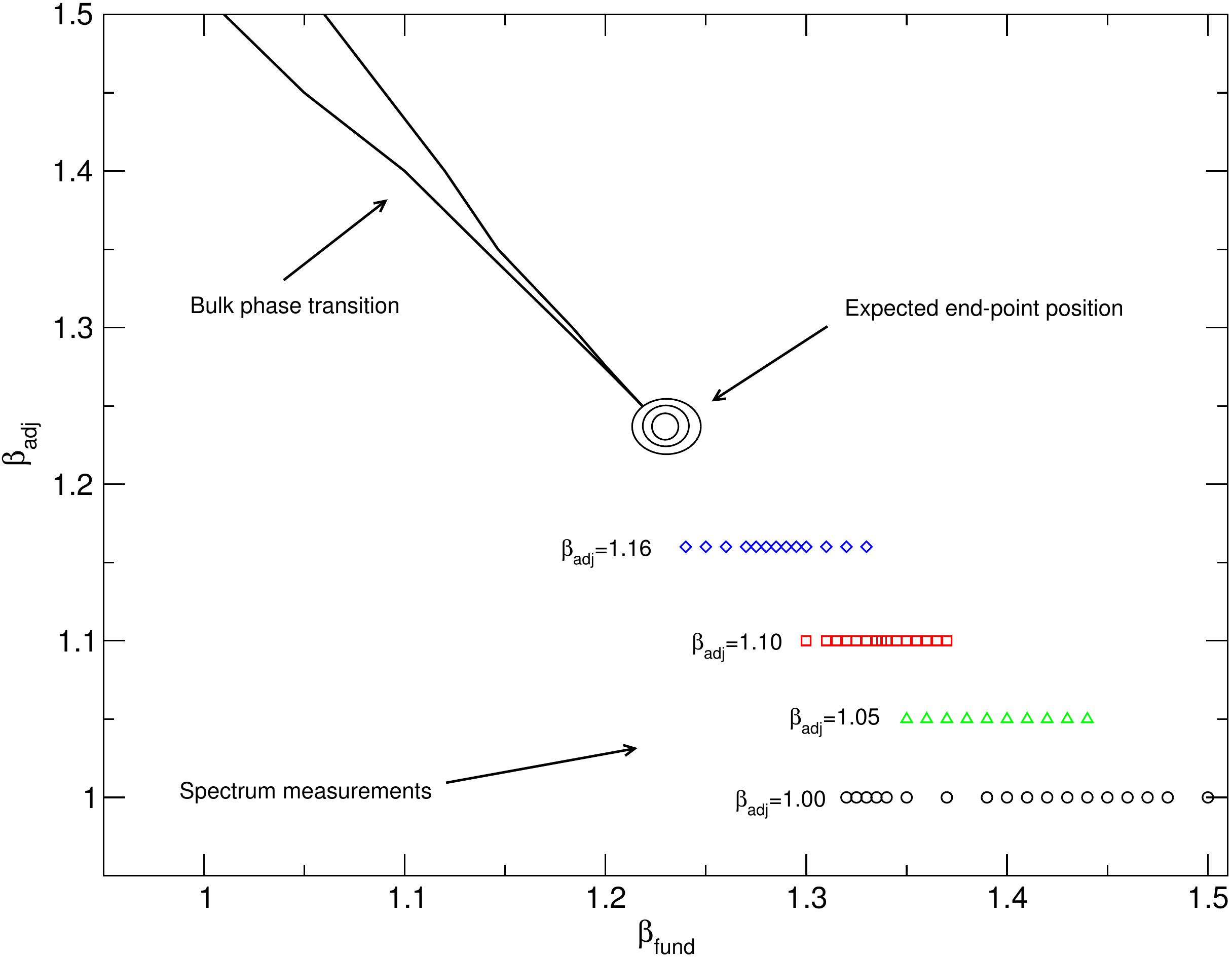}
  \caption{The location of the bulk phase transition is delimited by black
    lines, representing the extension of the hysteresis
    cycle. The approximate location of the bulk transition end--point
    is shown by the arrow. The coloured symbols show the points where
    the spectrum is investigated.}
  \label{fig:data-points}
\end{figure}
%%%%%%%%%%%%%%

\section{Spectrum measurements}
\label{sec:spectrum}

\begin{figure}[hb]
  \centering
  \begin{tabular}[h]{cc}
    \includegraphics[width=0.45\textwidth]{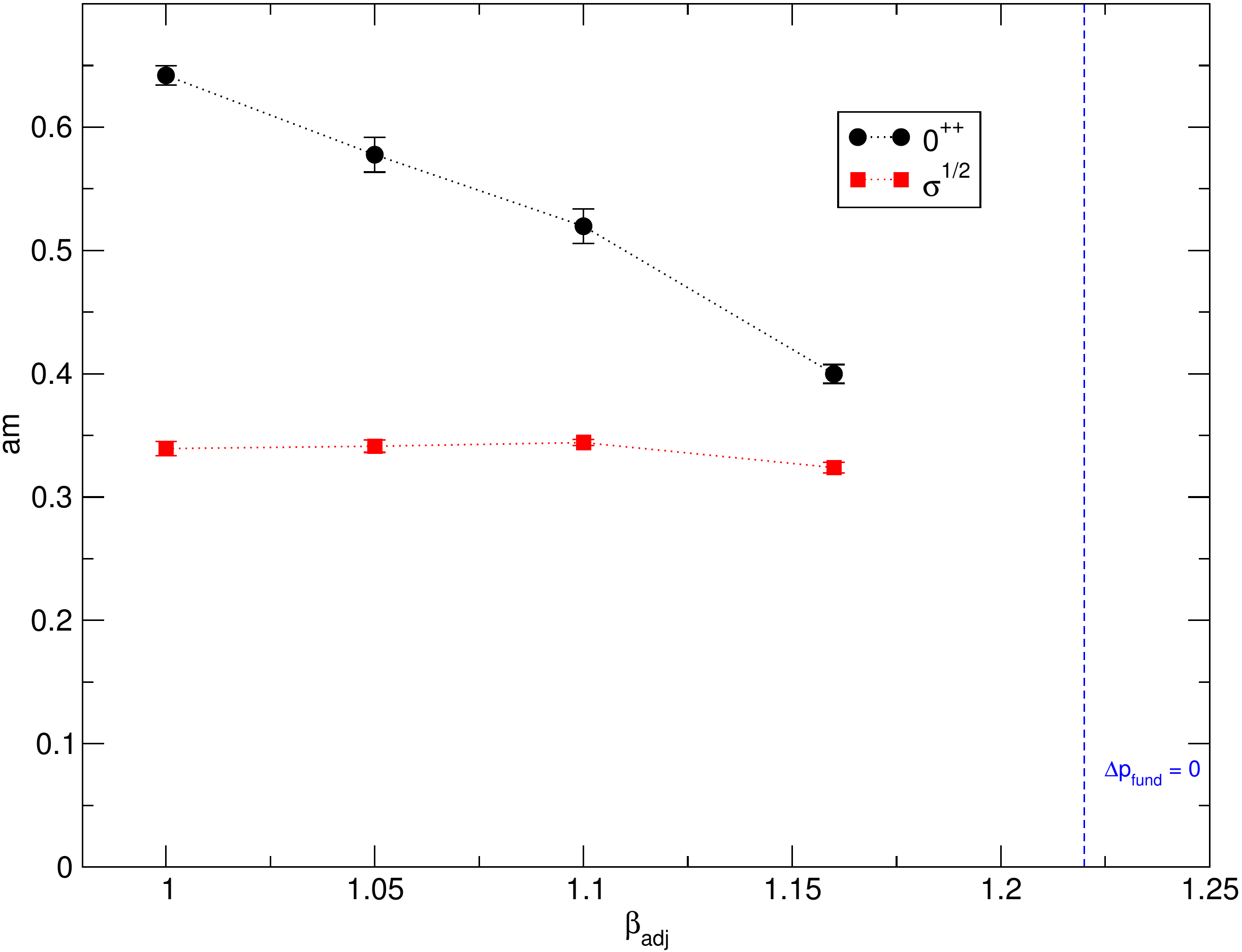} &
    \includegraphics[width=0.45\textwidth]{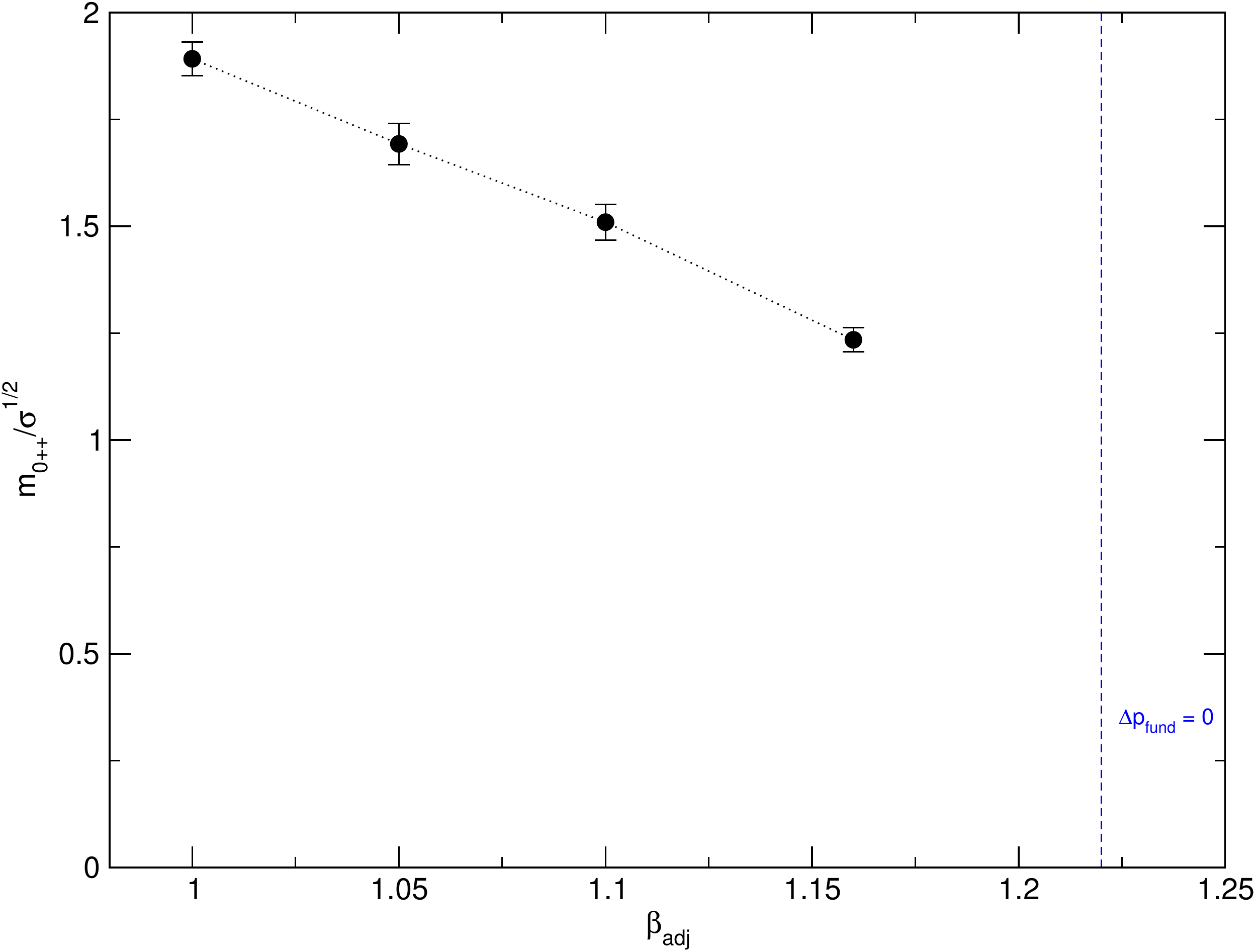} \\
  \end{tabular}
  \caption{(left) The scalar glueball mass and the square root of the
    string tension at the values ($\bfund$,$\badj$) defining the
    plaquette susceptibility's peak when approaching the end--point
    from below. (right) The ratio $\frac{m_{0^{++}}}{\sqrt{\sigma}}$
    on the same points. We highlighted the approximate location of the
    bulk transition end--point.}
  \label{fig:scaling-ratio}
\end{figure}

Let us first recall here that we do not want to precisely pin down the
end--point location, but rather to
identify its neighbourhood, where the spectrum of the theory should be
investigated. Our aim is to compare the scaling properties of the
spectrum when the bulk transition end--point is approached in a
controlled manner, with the ones of the model with $2$ adjoint
fermions. This will help us
clarify the still controversial nature of this end--point. In
Fig.~\ref{fig:data-points} a summary of our results concerning the
bulk phase transition line and its end--point is shown. In addition,
we indicate the points where we performed a detailed investigation of
the spectrum, as described in the following.\\
Our simulations with the fundamental--adjoint action are carried over
using a modified Metropolis algorithm~\cite{bazavov} which helps us
cope with the increasing autocorrelation times due to critical slowing
down when simulating closer to $\badj \approx 1.25$. % This algorithm has
% been shown to have excellent acceptance rate for all the point in
% Fig.~\ref{fig:data-points}. 
This allows us to obtain a large statistics
of independent gauge configurations even on large lattices when the
critical slowing down starts affecting the simulations. In particular,
we measure our observables on Monte Carlo histories of ${\mathcal
  O}(10000)$ configurations, each separated by ${\mathcal O}(100)$
modified--Metropolis updates of the SU(2) link matrices ($\badj$
values closer to the end--point have a larger number of intermediate
updates between measurements to reduce autocorrelations in our
ensembles). In the following, we show results at four different
values of the adjoint coupling $\badj = 1.00, \, 1.05, \, 1.10, \,
1.16$ and spanning a large range of $\bfund$ such that both regions
around the crossover are monitored. A sequence of five different
volumes is simulated for each $\badj$: $6^3\times12$, $10^3\times20$,
$16^3\times20$, $24^3\times32$ and $32^3\times32$, where the longer
temporal extent is used to % prevent large finite--temperature effects
% and to 
better identify effective mass plateaux.\\
We employ the variational procedure detailed in Ref.~\cite{biagio} to extract
the ground state mass and a few excitations of the spectrum in the
following channels: 
\begin{itemize}
\item {\bf String tension}: $a\sqrt{\sigma}$ is the lightest dynamical scale
  in a pure gauge theory and it is used to set the overall scale. We
  extract the string tension from correlators of long spatial Polyakov
  loops of length $L=aN_s$. The asymptotic large--time behaviour of
  these correlators is governed by the lightest torelon state whose
  mass $am_l$ can be used to obtain the string tension according to
  \begin{equation}
    \label{eq:sigma}
    am_l(N_s) \; = \; a^2\sigma N_s - \frac{\pi}{3N_s} -
    \frac{\pi^2}{18N_s^3}\frac{1}{a^2\sigma}
    \ .
  \end{equation}
  The validity of the above equation is checked {\em a posteriori} by
  comparing the extracted string tension with the one obtained using
  only the leading term $-\pi/(3N_s)$. Significant finite--size
  systematics are absent when $L\sqrt{\sigma} > 3$, which we satisfied
  in our simulations using large spatial volumes for the smallest values of
  $a\sqrt{\sigma}$.
\item {\bf Scalar glueball mass}: $am_{0^{++}}$ is the lightest glueball
  mass in the spectrum. Correlators of smeared spatial Wilson
  loops in the scalar representation of the cubic symmetry group are
  measured on each configuration.
\item {\bf Tensor glueball mass}: $am_{2^{++}}$ is the second lightest
  glueball and its mass is monitored to check whether its behaviour is
  the same as the scalar one. Having quantum numbers different from
  the vacuum, its scaling properties could be different in principle.
\end{itemize}
%%%%%%%%%%%%%%
\begin{figure}[ht]
  \centering
  \begin{tabular}[h]{cc}
    \includegraphics[width=0.45\textwidth]{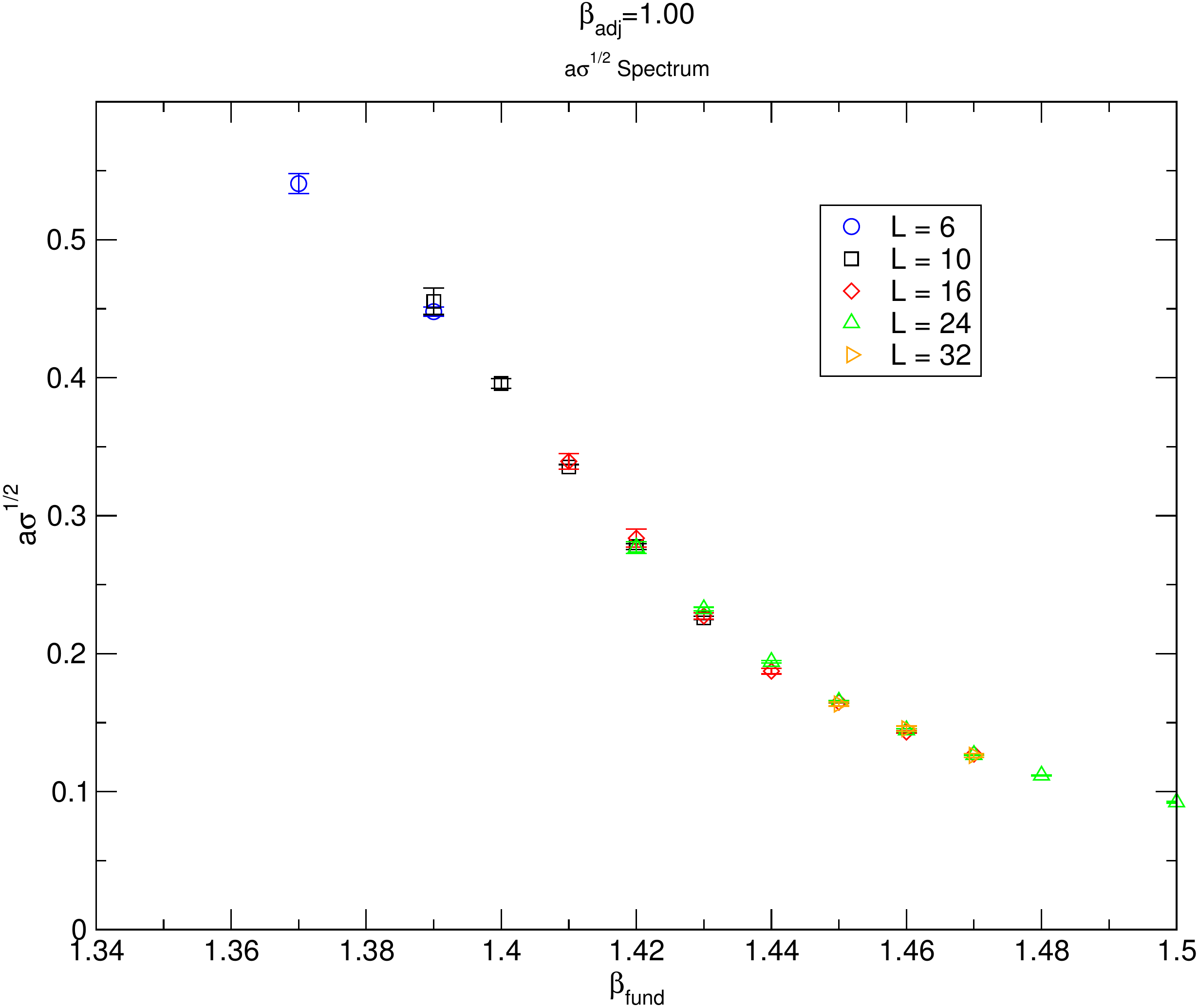} &
    \includegraphics[width=0.45\textwidth]{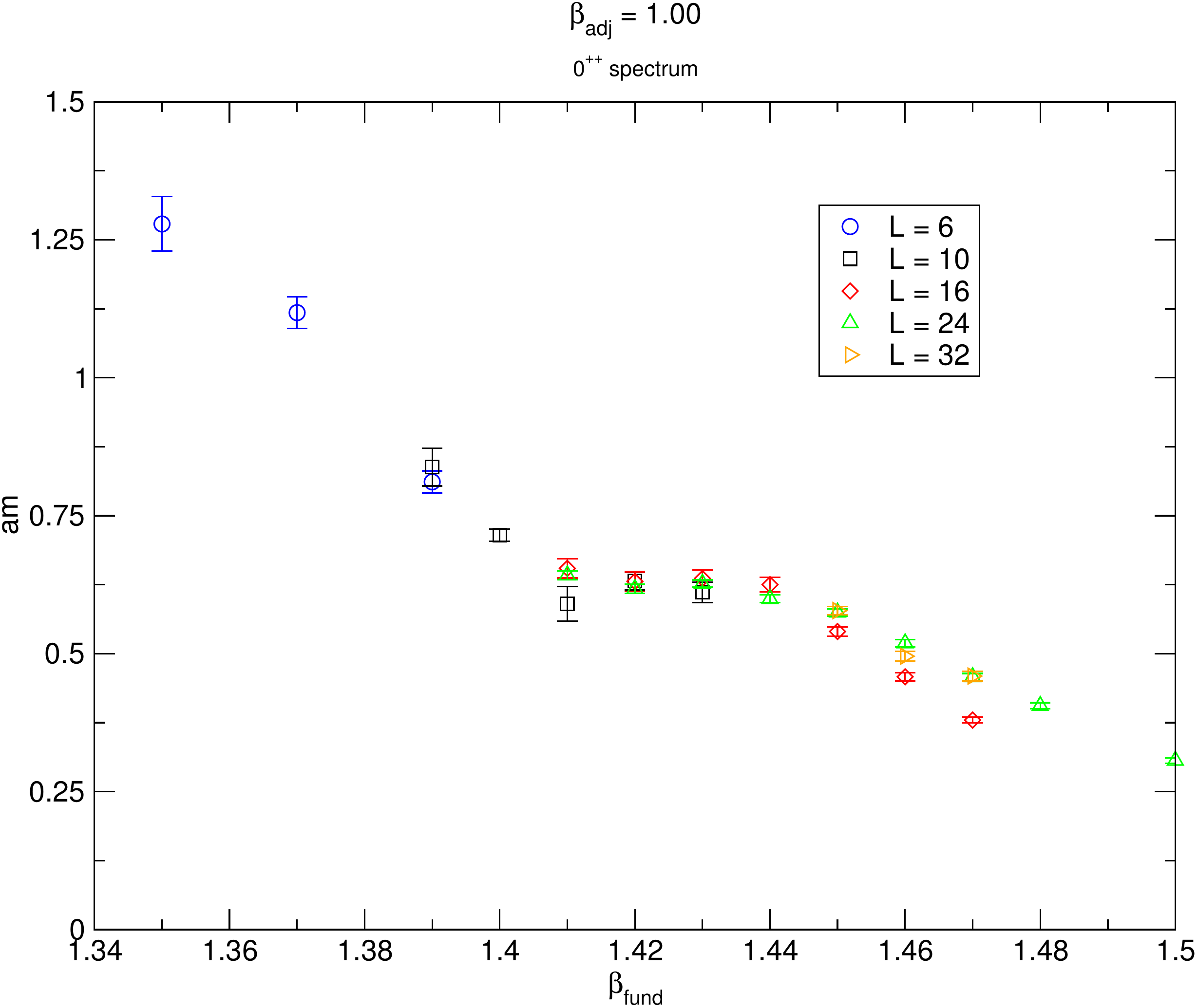} \\
  \end{tabular}
  \caption{Measured spectrum on different volumes at $\badj =
    1.00$. (left) The square root of the string tension in units of
    the lattice spacing as $\bfund$ increases towards the
    weak coupling region. (right) The mass of the lightest scalar
    glueball in units of the lattice spacing. Very good control over
    finite--size effects is obtained for both these observables.}
  \label{fig:spectrum-ba100}
\end{figure}
%%%%%%%%%%%%%%
In Fig.~\ref{fig:spectrum-ba100} we show the string tension
and scalar mass at $\badj=1.00$ and for a range of different $\bfund$
and spatial volumes $L^3$. The string tension decreases monotonically
when approaching the weak--coupling limit at large $\bfund$, whereas
the scalar glueball mass develops a short plateaux in the crossover
region % (cfr. Fig.~\ref{fig:plaquette-susc}(left))
before decreasing
again. In both cases we have a good control over finite--size effects,
with masses matching on at least two subsequent volumes for each
point. The largest finite--size effects are seen towards the weak
coupling, where the string tension becomes small. Not shown in the
plots is the behaviour of the tensor glueball which resembles the
string tension one, though with somewhat larger finite--volume
systematics.\\
%%%%%%%%%%%%%%
\begin{figure}[hb]
  \centering
  \begin{tabular}[h]{cc}
    \includegraphics[width=0.45\textwidth]{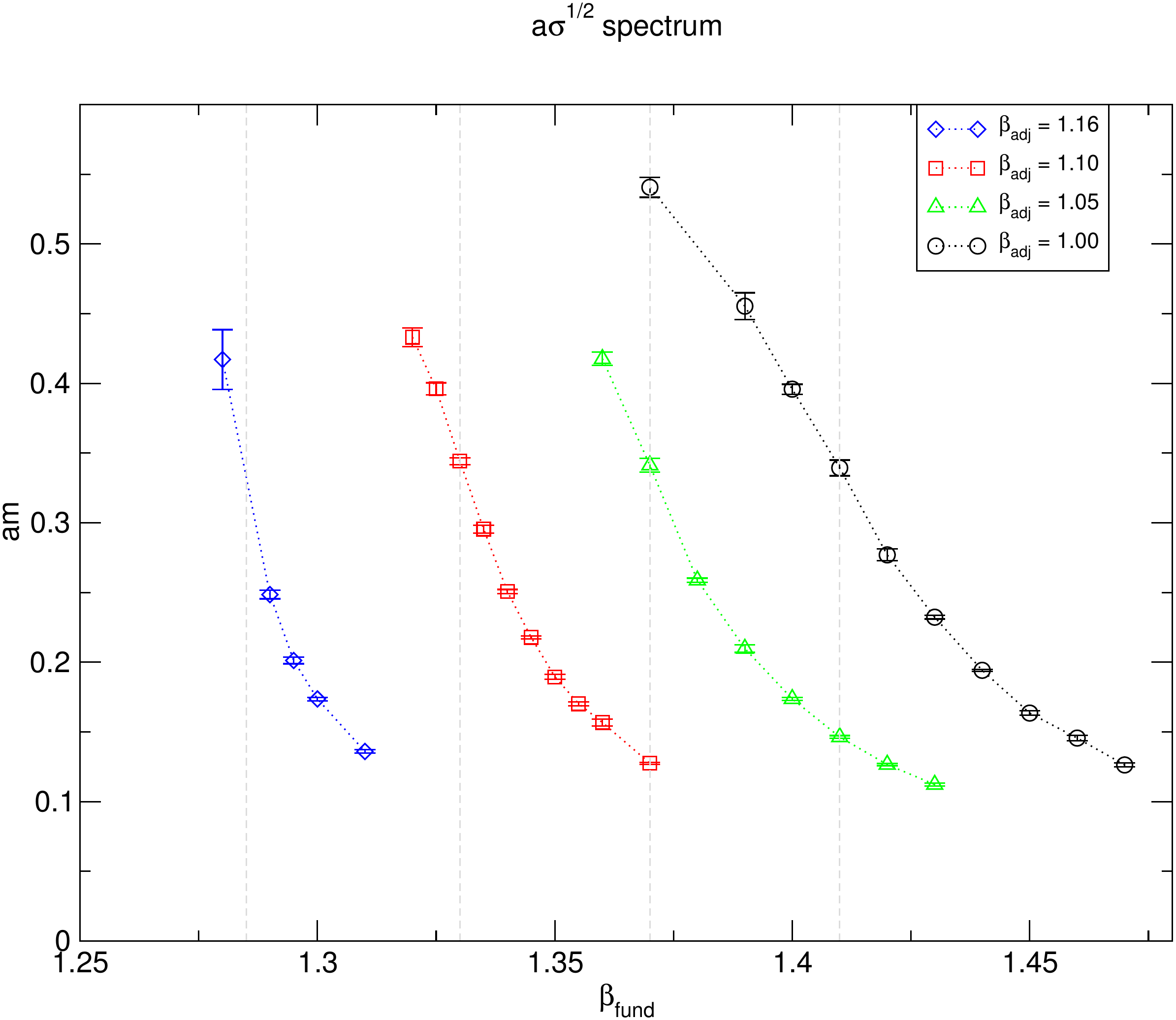} &
    \includegraphics[width=0.45\textwidth]{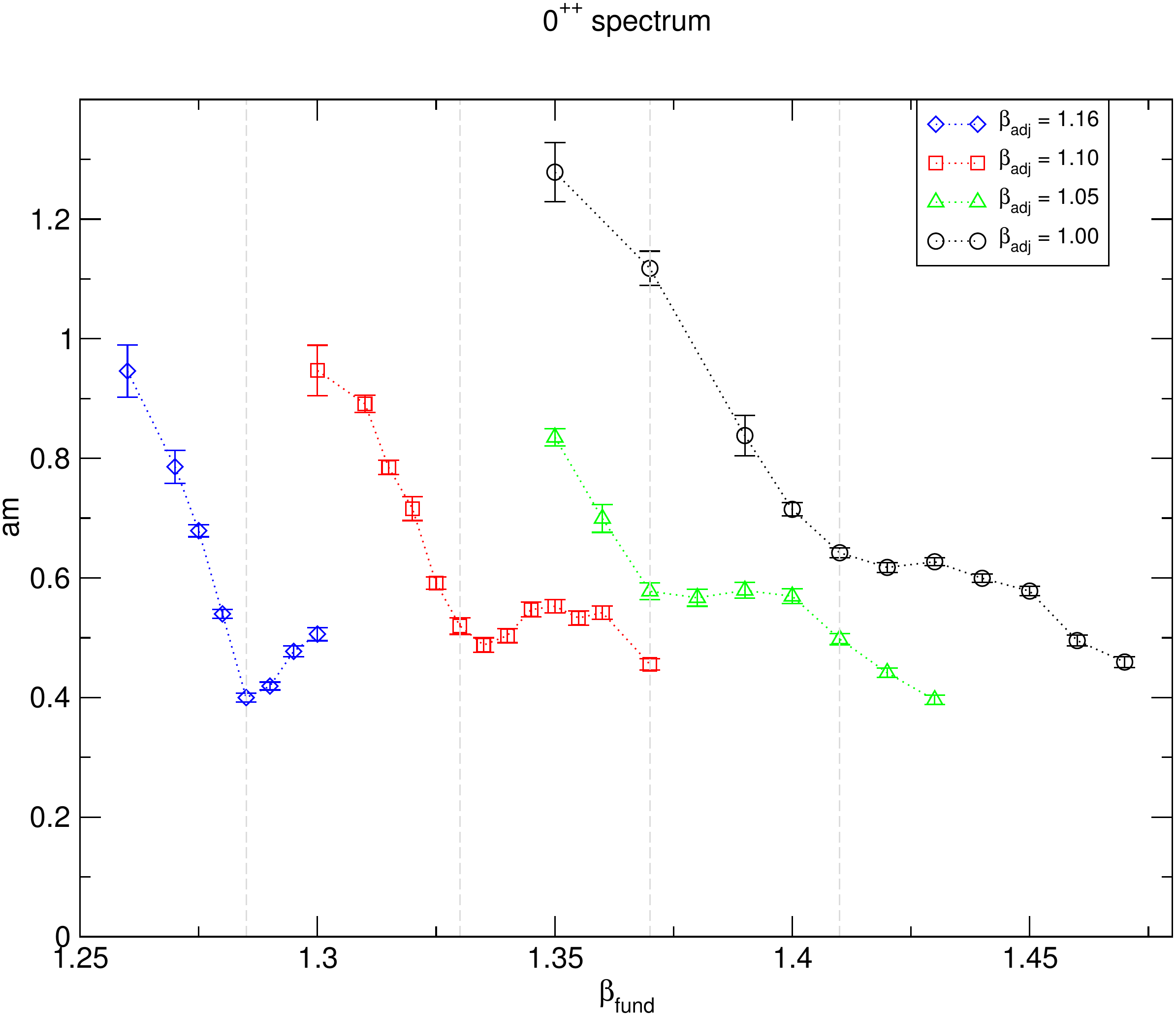} \\
  \end{tabular}
  \caption{(left) The infinite volume limit of $a\sqrt{\sigma}$ is
    plotted for $\badj$ closer and closer to the end--point
    location. (right) The scalar glueball mass in the infinite volume
    limit for all available $\badj$. The very different behaviour of
    the two observables is clear when $\badj$ is increased. Also shown
    are vertical dashed lines highlighting the location of the
    crossover region.}
  \label{fig:spectrum}
\end{figure}
%%%%%%%%%%%%%%
Given the large range of lattice volume used, we are able to reliably
estimate the infinite volume limit of $a\sqrt{\sigma}$ and
$am_{0^{++}}$ for all the $\badj$ values studied. However, for some of
these values we can not extrapolate $am_{2^{++}}$ at $L=\infty$ with
our current data. The comparison of the extracted infinite--volume
spectrum between different $\badj$ is shown in
Fig.~\ref{fig:spectrum}. To study the scaling of the observables along
a trajectory when approaching the bulk transition end--point, we
choose to follow the line in the phase diagram given by the peak of
the fundamental plaquette susceptibility. The location of such peak
at the four $\badj$ values investigated is highlighted by
vertical dashed lines in Fig.~\ref{fig:spectrum}. On those points, we
note that the string tension remains constant when approaching
the end--point (larger $\badj$), whereas the scalar glueball mass slightly
decreases: this suggest a non--constant ratio
$m_{0^{++}}/\sqrt{\sigma}$. A summary of this result is shown in
Fig.~\ref{fig:scaling-ratio}.
%%%%%%%%%%%%%%
%%%%%%%%%%%%%%

\section{Conclusions}
\label{sec:conclusions}

In this work we have studied a SU(2) pure gauge theory with a modified
lattice plaquette action. We added a coupling to plaquettes in
the adjoint representation of the gauge group. This lattice system is
known to have a bulk phase transition with an end--point relatively
close to the fundamental coupling axis. The nature of this end--point
is still controversial and we focused more on the region close to it,
but far from the bulk phase transition. Thanks to our improved gluonic
spectroscopic technique~\cite{biagio}, we measured the string tension,
the scalar glueball mass and the tensor one, aiming at studying their
scaling properties when the end--point is approached. This is the
first study of the gluonic spectrum in this model. Therefore we
carefully checked for finite--size systematics and tried to reduce
autocorrelation effects on our observables. The extrapolated
infinite--volume spectrum shows a non--constant
$m_{0^{++}}/\sqrt{\sigma}$ ratio when approaching the end--point
in a controlled manner. This seems in contrast with the infrared
dynamics of the SU(2) theory with 2 adjoint fermions, where such a
ratio is driven by a conformal fixed point and is consistent with the
continuum SU(2) Yang--Mills value $m_{0^{++}}/\sqrt{\sigma} \sim
3.7$. The
results presented here should be considered as preliminary and might
be still too far from the basin of attraction of the end--point. More
detailed analysis and discussions will be presented in a forthcoming
publication~\cite{mypaper}.

%-----------------------BIBLIOGRAPHY----------------------------

\end{document}